\documentclass[twocolumn,prb]{revtex4}
\usepackage{amsfonts}
\usepackage[T1]{fontenc}
\usepackage{amsmath,amsbsy,amssymb,graphicx}
\usepackage{times}

\let\mathbf=\boldsymbol
\begin{document}

\title{Systematic construction of square-root topological insulators and
superconductors}
\author{Motohiko Ezawa}
\affiliation{Department of Applied Physics, University of Tokyo, Hongo 7-3-1, 113-8656,
Japan}

\begin{abstract}
We propose a general scheme to construct a Hamiltonian $H_{\text{root}}$
describing a square root of an original Hamiltonian $H_{\text{original}}$
based on the graph theory. The square-root Hamiltonian is defined on the
subdivided graph of the original graph of $H_{\text{original}}$, where the
subdivided graph is obtained by putting one vertex on each link in the
original graph. When $H_{\text{original}}$ describes a topological system,
there emerge in-gap edge states at non-zero energy in the spectrum of $H_{\text{root}}$, 
which are the inherence of the topological edge states at
zero energy in $H_{\text{original}}$. In this case, $H_{\text{root}}$
describes a square-root topological insulator or superconductor. Typical
examples are square roots of the Su-Schrieffer-Heeger (SSH) model, the
Kitaev topological superconductor model and the Haldane model. Our scheme is
also applicable to non-Hermitian topological systems, where we study an
example of a nonreciprocal non-Hermitian SSH model.
\end{abstract}

\maketitle

\textbf{Introduction:} Topological insulators and superconductors are among
the most studied fields in condensed matter physics in this decade\cite%
{Hasan,Qi}. They are characterized by the bulk-edge correspondence, i.e., by
the emergence of topological edge states although the bulk is gapped.

Recently, a square-root topological insulator is proposed\cite{Ark,Kremer}.
Its notion has also been generalized to square-root higher-order topological
insulators\cite{Hatsugai}. They are characterized by the emergence of in-gap
edge states at nonzero energy, which are the inherence of the topological
edge states at zero energy of the original Hamiltonian\cite{Ark,Kremer}.

\begin{figure}[t]
\centerline{\includegraphics[width=0.49\textwidth]{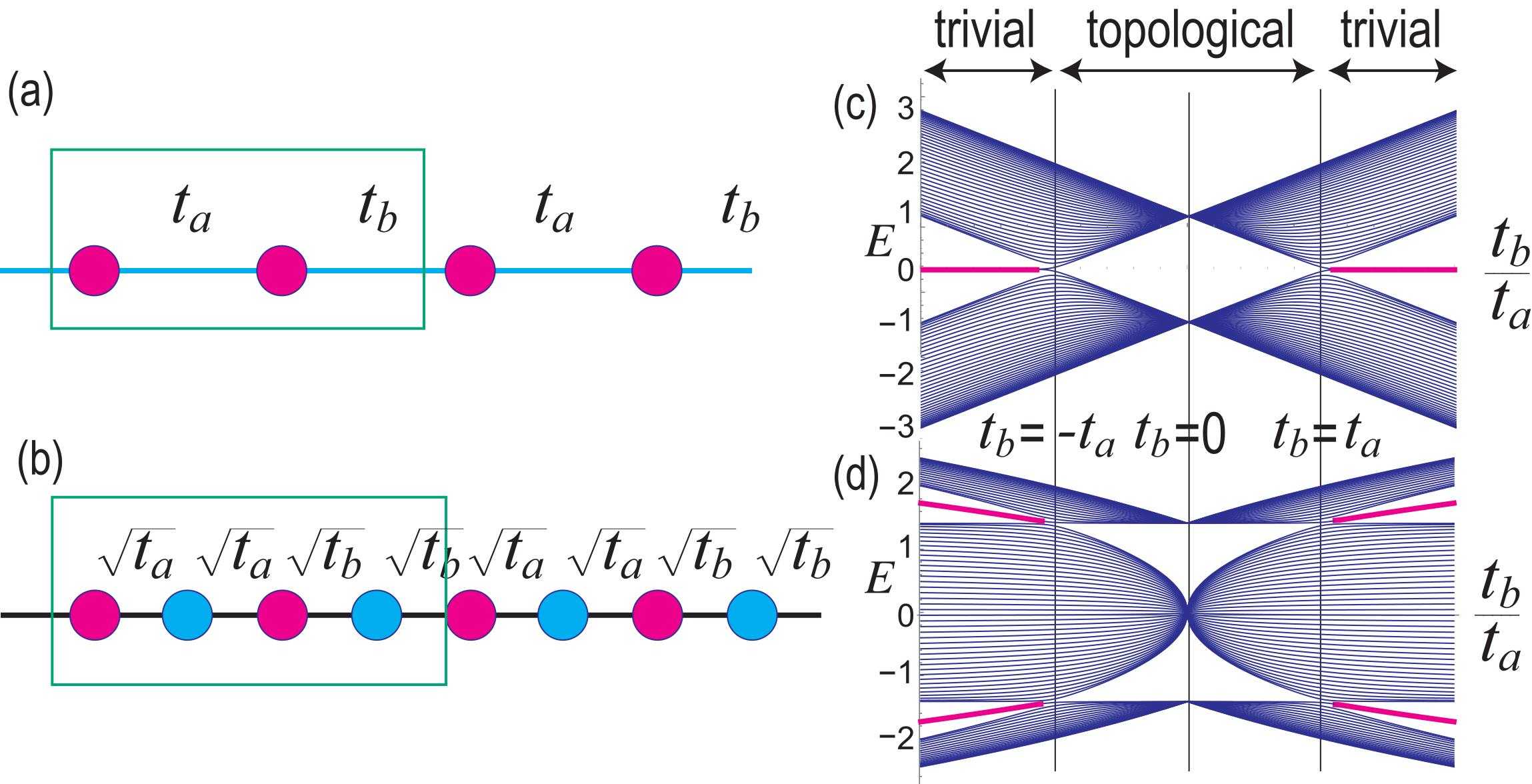}}
\caption{Illustration of (a) the graph and (b) the subdivided graph for the
SSH model $H_{\text{SSH}}$. The green rectangles represent the unit cells.
(c) Energy spectrum of $H_{\text{SSH}}$ and (d) $H_{\text{root}}$ in unit of 
$t_a$ as a function of $t_{b}/t_{a}$. The topological edge states are marked
in magenta. In-gap edge states are on the curve 
$E=\pm \protect\sqrt{\left\vert t_{a}\right\vert +\left\vert t_{b}\right\vert }$ for $H_{\text{root}}$. }
\label{FigSqSSH}
\end{figure}

\begin{figure}[t]
\centerline{\includegraphics[width=0.49\textwidth]{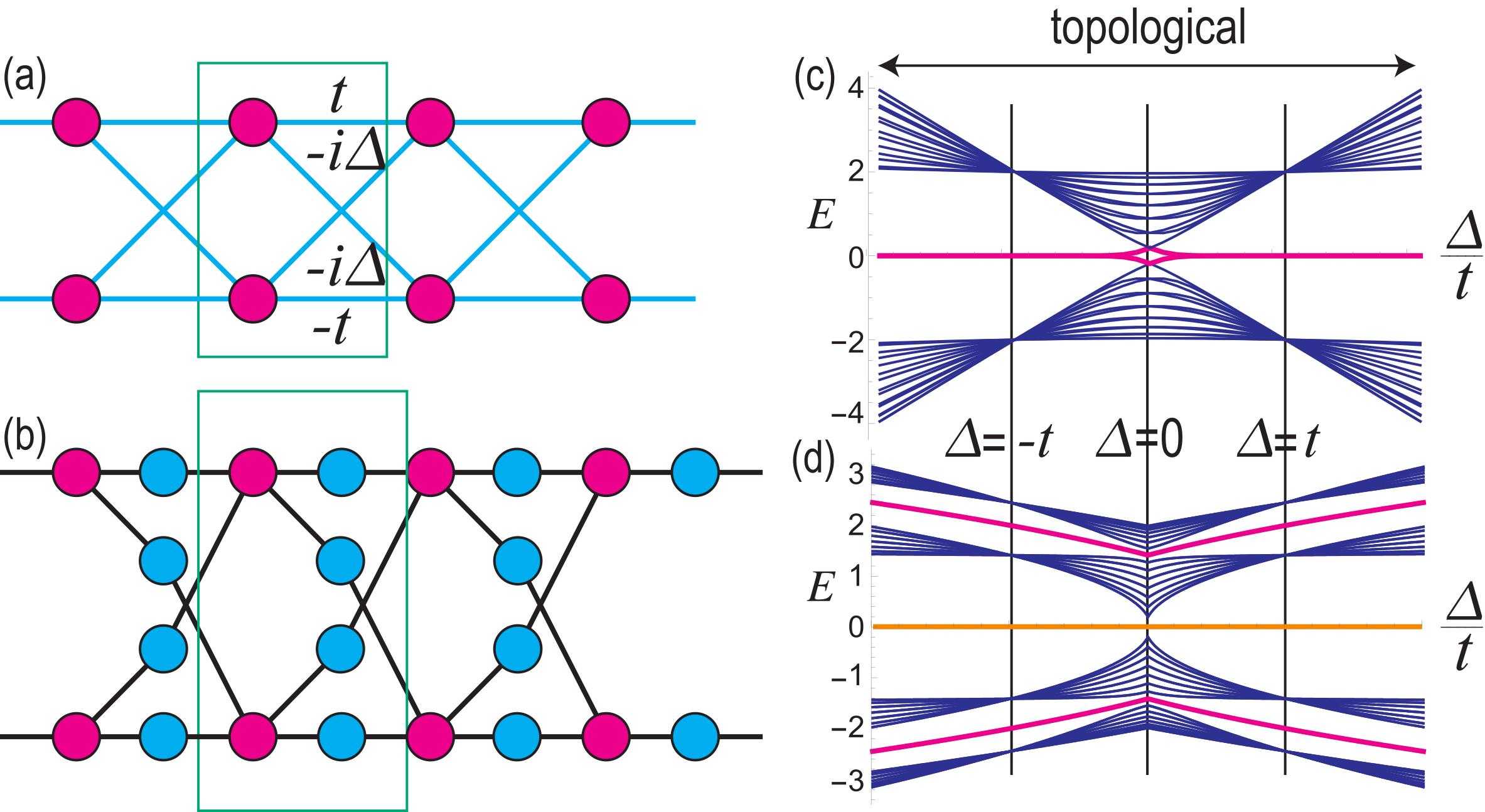}}
\caption{Illustration of (a) the graph and (b) the subdivided graph for the
Kitaev model $H_{\text{Kitaev}}$. (c) Energy spectrum of $H_{\text{Kitaev}}$
and (d) $H_{\text{root}}$ in unit of $t$ as a function of $\Delta /t$. The
topological edge states are marked in magenta. In-gap states are on the
curve $E=\pm \protect\sqrt{2\left\vert t\right\vert +2\left\vert \Delta
\right\vert }$ for $H_{\text{root}}$. Lieb perfect-flat bulk-bands are
marked in orange.}
\label{FigSqKitaev}
\end{figure}

In this paper, we propose a general scheme to construct square-root
topological insulators and superconductors from ordinary topological
insulators and superconductors based on the graph theory. There is
one-to-one correspondence between a tight-binding Hamiltonian and a weighted
graph. A graph is composed of vertices and links. We can construct a new
graph by introducing one vertex on each link, which we refer to as the
subdivided graph\cite{EzawaSch,EzawaUniv}. We call the original graph the
parent graph in contrast to the subdivided graph. Any subdivided graph is
bipartite because it contains original vertices and newly added vertices.
Examples are shown in Figs.\ref{FigSqSSH}, \ref{FigSqKitaev} and \ref%
{FigSqHaldane}, where original (new) vertices are shown in magenta (cyan).

\begin{figure}[t]
\centerline{\includegraphics[width=0.49\textwidth]{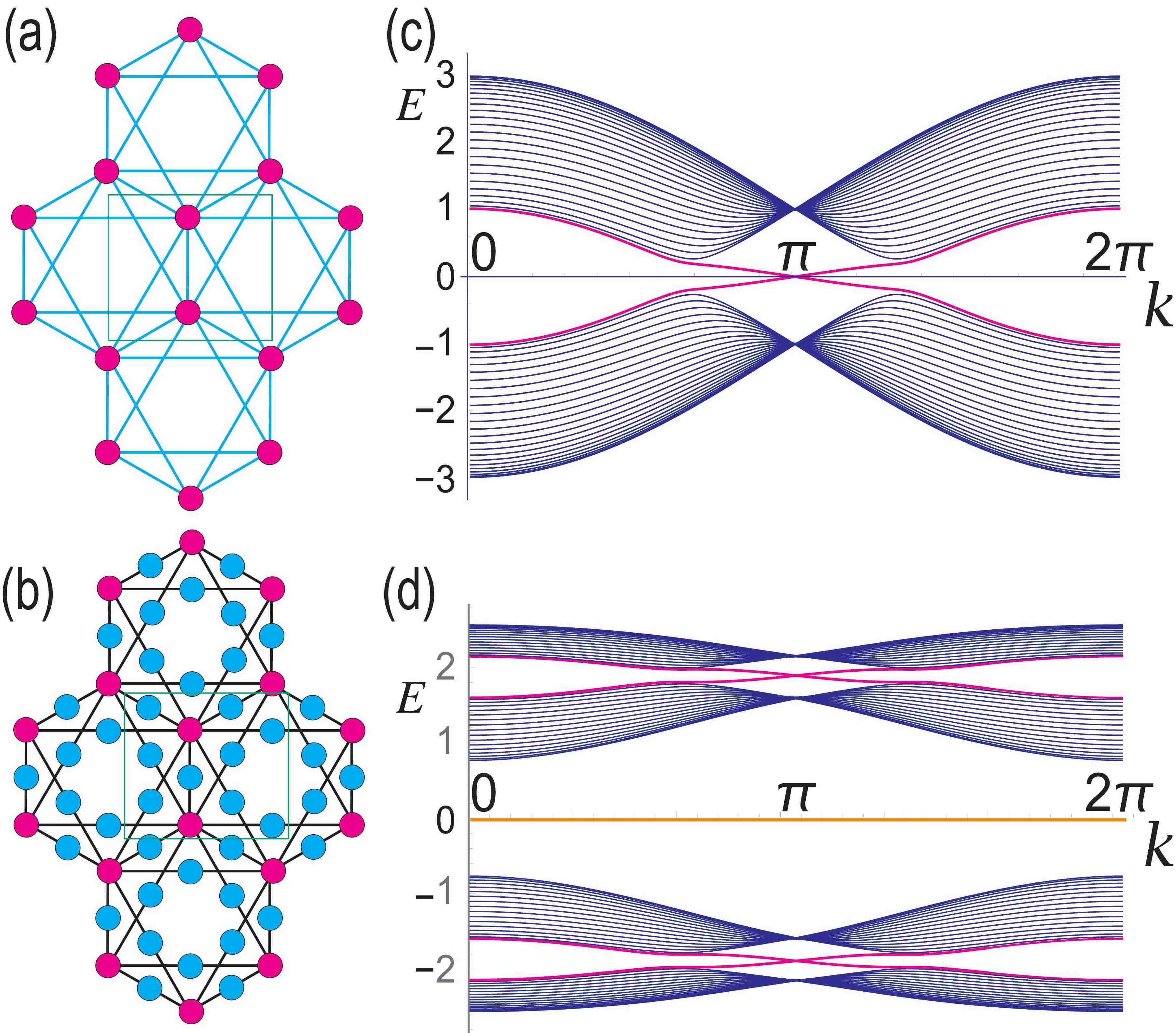}}
\caption{Illustration of (a) the graph and (b) the subdivided graph for the
Haldane model $H_{\text{Haldane}}$. (c) Energy spectrum of $H_{\text{Haldane}}$ 
and (d) $H_{\text{root}}$ in unit of $t$ as a function of the momentum $k$. 
The chiral edge states are marked in magenta. Lieb perfect-flat bulk-bands
are marked in orange. We have set $\protect\lambda =0.2t /(3\protect\sqrt{3}) $.}
\label{FigSqHaldane}
\end{figure}

We denote the Hamiltonian constructed on the subdivided graph as $H_{\text{root}}$. 
We then find $(H_{\text{root}})^{2}=H_{\text{par}}\oplus H_{\text{res}}$, 
where $H_{\text{par}}$ is identical to the original Hamiltonian $H_{\text{original}}$ 
up to an additive constant interpreted as a self-energy.
We call $H_{\text{par}}$ and $H_{\text{res}}$ the parent and residual
Hamiltonians, respectively. When $H_{\text{original}}$ describes a
topological system, it contains zero-energy topological boundary states,
producing the corresponding boundary states at nonzero energy in $H_{\text{root}}$. 
Furthermore, zero-energy perfect-flat bulk-bands may emerge in $H_{\text{root}}$ 
as a bipartite property according to the Lieb theorem: See
orange lines in Figs.\ref{FigSqKitaev} and \ref{FigSqHaldane}. Because the
eigenvalues are shown to be identical between $H_{\text{par}}$ and 
$H_{\text{res}}$ except for zero-energy states in $H_{\text{res}}$, $H_{\text{root}}$
is interpreted as the square-root Hamiltonian of $H_{\text{original}}$. We
can use the same topological index between $H_{\text{root}}$ and 
$H_{\text{original}}$ since the eigenvectors are identical between them. Indeed, the
region of the in-gap edge states in $H_{\text{root}}$ is precisely the same
as that of the zero-energy edge states in $H_{\text{original}}$.

We present explicit examples of the Su-Schrieffer-Heeger (SSH), the Kitaev 
$p $-wave topological superconductor model and the Haldane honeycomb model.
Furthermore, our results are applicable to non-Hermitian systems, where we
demonstrate an example of nonreciprocal non-Hermitian SSH model.

\textbf{Square-root Hamiltonian:} It is impossible to construct a local
hopping model only by taking a square root of the Hamiltonian matrix. A
simple example is given by the SSH model, 
\begin{equation}
H_{\text{SSH}}=\left( 
\begin{array}{cc}
0 & t_{a}+t_{b}e^{-ik} \\ 
t_{a}+t_{b}e^{ik} & 0%
\end{array}%
\right) .  \label{HamilSSH}
\end{equation}%
A square root of the model is given by\cite{NoteSM1}%
\begin{equation}
\sqrt{H_{\text{SSH}}}=\left( 
\begin{array}{cc}
0 & \frac{\sqrt{E\left( k\right) }}{t_{a}+t_{b}e^{-ik}} \\ 
\frac{t_{a}+t_{b}e^{-ik}}{\sqrt{E\left( k\right) }} & 0%
\end{array}%
\right) ,
\end{equation}%
with $E\left( k\right) =\sqrt{t_{a}^{2}+t_{b}^{2}+2t_{a}t_{b}\cos k}$. This
is an infinite-range hopping model.

Here we recall the Dirac idea to take a square root of the Klein-Gordon
equation. He obtained the Dirac equation by introducing a matrix degree of
freedom. The Dirac equation has various intriguing properties such as
chirality and the index theorem, which are absent in the Klein-Gordon
equation.

We propose to take a square root of a Hamiltonian by increasing a matrix
degree of freedom as follows: 1) We first write down a graph representation
of the adjacency matrix of the original Hamiltonian $H_{\text{original}}$.
2) We construct a subdivided graph from the original graph. 3) We construct
a Hamiltonian $H_{\text{root}}$ on the subdivided graph. Then, we obtain 
$(H_{\text{root}})^{2}=H_{\text{par}}\oplus H_{\text{res}}$, 
where $H_{\text{par}}$ is identical to the original Hamiltonian $H_{\text{original}}$ up to
an additive constant, provided the hopping parameter is taken to be $\sqrt{t}
$ in $H_{\text{root}}$ corresponding to the hopping parameter $t$ in 
$H_{\text{original}}$. The square-root Hamiltonian is given by $H_{\text{root}}$.

We start with a Hamiltonian $H_{\text{original}}$ where a unit cell contains 
$N$ sites connected by $M$ hoppings. We consider a Hamiltonian on the
subdivided graph, which is given by%
\begin{equation}
H_{\text{root}}=\left( 
\begin{array}{cc}
O_{N\times N} & H_{N\times M}^{\text{left}} \\ 
H_{M\times N}^{\text{right}} & O_{M\times M}%
\end{array}%
\right) .  \label{HamilSR}
\end{equation}%
It is required that $H_{M\times N}^{\text{right}}=(H_{N\times M}^{\text{left}%
})^{\dagger }$, when $H_{\text{root}}$ is Hermitian. We have%
\begin{equation}
(H_{\text{root}})^{2}=\left( 
\begin{array}{cc}
H_{\text{par}} & 0 \\ 
0 & H_{\text{res}}%
\end{array}%
\right) =H_{\text{par}}\oplus H_{\text{res}},  \label{ParRes}
\end{equation}%
where 
\begin{equation}
H_{\text{par}}\equiv H_{N\times M}^{\text{left}}H_{M\times N}^{\text{right}},
\quad H_{\text{res}}\equiv H_{M\times N}^{\text{right}}H_{N\times M}^{\text{left}}.
\end{equation}%
Thus the square of the Hamiltonian, $(H_{\text{root}})^{2}$, is decomposed
into a direct sum of $H_{\text{par}}$ and $H_{\text{res}}$, which are the
parent and residual Hamiltonians defined on the parent and residual graphs.

In general, $H_{\text{par}}$ is identical to $H_{\text{original}}$ up to a
constant term because both of them are constructed on the same graph,%
\begin{equation}
H_{\text{par}}=C+H_{\text{original}},  \label{Const}
\end{equation}%
where $C$ is a positive constant obtained by calculating $(H_{\text{root}})^{2}$. 
This constant term can be interpreted as a self energy, as in the
case of the second-order perturbation theory. The zero-energy topological
edge states in $H_{\text{original}}$ are transformed into the in-gap
boundary states at non-zero energy $\pm \sqrt{C}$ in $H_{\text{root}}$.

The bipartite Hamiltonian has chiral symmetry, $\{\gamma ,H_{\text{root}}\}=0 $, with the chiral operator defined by%
\begin{equation}
\gamma =\left( 
\begin{array}{cc}
I_{N\times N} & O_{N\times M} \\ 
O_{M\times N} & -I_{M\times M}%
\end{array}%
\right) .
\end{equation}%
In general, we have $M>N$. According to the Lieb theorem\cite{Lieb}, there
are $\left\vert M-N\right\vert $ zero-energy states constituting
perfect-flat bulk bands.

It is known\cite{Das,Tsune,NoteSM2} that the eigenvalues are identical
between $H_{\text{par}}$ and $H_{\text{res}}$\ except for these zero-energy
states in $H_{\text{res}}$ and that they are non-negative. Namely, when we
set%
\begin{equation}
H_{\text{par}}|\psi _{j}^{\text{par}}\rangle =\varepsilon _{j}|\psi _{j}^{\text{par}}\rangle ,
\quad H_{\text{res}}|\psi _{j}^{\text{res}}\rangle
=\varepsilon _{j}^{\text{res}}|\psi _{j}^{\text{res}}\rangle ,
\end{equation}%
we obtain $\{\varepsilon _{j}^{\text{res}}\}=\{\varepsilon _{1},\cdots
\varepsilon _{N},0,\cdots ,0\}$ and $\varepsilon _{j}\geq 0$. It follows
from (\ref{Const}) that the eigenvectors of $H_{\text{par}}$ and 
$H_{\text{original}}$ are the same, $H_{\text{original}}|\psi _{j}^{\text{par}}\rangle
=(\varepsilon _{j}-C)|\psi _{j}^{\text{par}}\rangle $. Furthermore, the
eigenvectors $|\psi _{j}^{\text{res}}\rangle $ of $H_{\text{res}}$ are
obtained from those of $H_{\text{par}}$ as\cite{NoteSM2}%
\begin{equation}
|\psi _{j}^{\text{res}}\rangle \equiv \sum_{k}(H_{M\times N}^{\text{right}})_{jk}|\psi _{k}^{\text{par}}\rangle .
\end{equation}%
The eigenvectors of $(H_{\text{root}})^{2}$ are given by 
\begin{equation}
|\psi _{j}^{\text{root}}\rangle =\{\psi _{1}^{\text{par}},\cdots ,\psi _{N}^{\text{par}},
\psi _{1}^{\text{res}},\cdots ,\psi _{M}^{\text{res}}\}.
\end{equation}%
When the Hamiltonian $H_{\text{root}}$\ is diagonalized by a unitary
transformation $U$ as $U^{\dagger }H_{\text{root}}U=H_{\text{root}}^{\text{D}}$, 
the Hamiltonian $(H_{\text{root}})^{2}$ is also diagonalized by the same
unitary transformation as $U^{\dagger }(H_{\text{root}})^{2}U=(H_{\text{root}}^{\text{D}})^{2}$. 
Then, the eigenvalues of $H_{\text{root}}$ are obtained
just by taking a square root of them with the same eigenvectors,%
\begin{equation}
H_{\text{root}}|\psi _{j}^{\text{root}}\rangle =\varepsilon _{j}^{\text{root}}|\psi _{j}^{\text{root}}\rangle ,
\end{equation}%
where 
\begin{equation}
\varepsilon _{j}^{\text{root}}=\{\pm \sqrt{\varepsilon _{1}},\cdots ,\pm 
\sqrt{\varepsilon _{N}};\pm \sqrt{\varepsilon _{1}},\cdots ,\pm \sqrt{\varepsilon _{N}},0,\cdots ,0\}.
\end{equation}%
Because of this property, $H_{\text{root}}$ is interpreted as the
square-root Hamiltonian of $H_{\text{original}}$.

An important observation is that the topolgocal properties are identical
between $H_{\text{root}}$ and $H_{\text{original}}$ since the eigenvectors
are the same. Correspondingly, the topological indices are the same.

\textbf{Square-root SSH model:} For the first example, we analyze the SSH
model (\ref{HamilSSH}). The spectrum contains the zero-energy topological
edge states as in Fig.\ref{FigSqSSH}(c). The graph of the SSH model is a
simple one-dimensional graph containing two vertices in the unit cell [Fig.\ref{FigSqSSH}(a)]. 
The corresponding subdivided graph is a one-dimensional
graph containing four vertices in the unit cell [Fig.\ref{FigSqSSH}(b)]. The
square-root Hamiltonian $H_{\text{root}}$ is given by (\ref{HamilSR}) with $(N,M)=(2,2)$, and%
\begin{equation}
H_{2\times 2}^{\text{left}}=\left( 
\begin{array}{cc}
\sqrt{t_{a}} & \sqrt{t_{b}}e^{-ik} \\ 
\sqrt{t_{a}} & \sqrt{t_{b}}%
\end{array}%
\right) .
\end{equation}%
It is straightforward to derive $(H_{\text{root}})^{2}=H_{\text{par}}\oplus
H_{\text{res}}$ with%
\begin{align}
H_{\text{par}} &=\left\vert t_{a}\right\vert +\left\vert t_{b}\right\vert
+H_{\text{SSH}}, \\
H_{\text{res}} &=\left( 
\begin{array}{cc}
2t_{a} & \sqrt{t_{a}t_{b}}\left( 1+e^{-ik}\right) \\ 
\sqrt{t_{a}t_{b}}\left( 1+e^{ik}\right) & 2t_{b}%
\end{array}%
\right) ,
\end{align}%
where $H_{\text{res}}$ is the Rice-Mele model. In-gap edge states appear at 
$E=\pm \sqrt{\left\vert t_{a}\right\vert +\left\vert t_{b}\right\vert }$ for 
$\left\vert t_{b}\right\vert >\left\vert t_{a}\right\vert $, as illustrated
in Fig.\ref{FigSqSSH}(d), whose origin is the topological zero-energy states
in the SSH model [Fig.\ref{FigSqSSH}(c)].

\textbf{Square-root Kitaev topological superconductor:} The next example is
a square root of the Kitaev $p$-wave topological superconductor model
defined by\cite{Kitaev,Alicea,Flen,Beenakker}%
\begin{equation}
H_{\text{Kitaev}}=\left( 2t\cos k-\mu \right) \sigma _{z}+2\Delta \sigma
_{x}\sin k.
\end{equation}%
The spectrum contains the zero-energy topological edge states as in Fig.\ref{FigSqKitaev}(c).
The corresponding graph and subdivided graph are shown in
Fig.\ref{FigSqKitaev}(a) and (b). The square-root Hamiltonian $H_{\text{root}}$ 
is given the Hamiltonian (\ref{HamilSR}) with $(N,M)=(2,4)$, and%
\begin{equation}
H_{2\times 4}^{\text{left}}=\left( 
\begin{array}{cccc}
\sqrt{t}\left( 1+e^{-ik}\right) & 0 & \Delta ^{\prime } & \Delta ^{\prime
\ast }e^{-ik} \\ 
0 & i\sqrt{t}\left( 1-e^{-ik}\right) & \Delta ^{\prime \ast }e^{-ik} & 
\Delta ^{\prime }%
\end{array}%
\right) ,
\end{equation}%
where $\Delta ^{\prime }=e^{-i\pi /4}\sqrt{\Delta }$.

We calculate $(H_{\text{root}})^{2}=H_{\text{par}}\oplus H_{\text{res}}$.
The parent Hamiltonian $H_{\text{par}}$ is found to be the Kitaev Hamiltonian%
\cite{Kitaev,Alicea,Flen,Beenakker} with $\mu =0$ and the addition of a
constant term $2\left\vert t\right\vert +2\left\vert \Delta \right\vert $.
In-gap edge states appear in $H_{\text{root}}$ at $E=\pm \sqrt{2\left\vert
t\right\vert +2\left\vert \Delta \right\vert }$ as in Fig.\ref{FigSqKitaev}(d), 
which are transformed from the zero-energy topological states in the
Kitaev model[Fig.\ref{FigSqKitaev}(c)]. Furthermore, there are perfect-flat
bulk-bands at zero energy in $H_{\text{root}}$ due to the Lieb theorem\cite{Lieb} with $\left\vert M-N\right\vert =2$.

\textbf{Square-root Haldane model:} We next study a square root of the
Haldane model. The Hamiltonian is defined on the graph in Fig.\ref{FigSqHaldane}(a) and given by%
\begin{align}
H_{\text{Haldane}} &=2\lambda \left( 2\sin \frac{k_{x}}{2}\cos \frac{\sqrt{3}%
k_{y}}{2}-\sin k_{x}\right) \sigma _{z}  \notag \\
&+t\left( 1+\cos \frac{\sqrt{3}k_{x}}{2}\cos \frac{k_{y}}{2}\right) \sigma
_{x}  \notag \\
&+t\left( \cos \frac{\sqrt{3}k_{x}}{2}\sin \frac{k_{y}}{2}\right) \sigma
_{y}.
\end{align}%
The spectrum in nanoribbon geometry contains chiral edge states as in 
Fig.\ref{FigSqHaldane}(c). The subdivided graph of the honeycomb graph is shown
in Fig.\ref{FigSqHaldane}(b). The square-root Hamiltonian $H_{\text{root}}$
is given by the Hamiltonian (\ref{HamilSR}) with $(N,M)=(2,9)$ and 
$H_{2\times 9}=\{a_{ij}\}$, where $a_{11}=\sqrt{t}$, $a_{12}=\sqrt{t}e^{i\mathbf{k}\cdot \mathbf{a}_{2}}$, 
$a_{13}=\sqrt{t}e^{-i\mathbf{k}\cdot 
\mathbf{a}_{1}}$, $a_{14}=\lambda +\lambda ^{\ast }e^{i\mathbf{k}\cdot 
\mathbf{a}_{2}}$, $a_{15}=\lambda ^{\ast }+\lambda e^{-i\mathbf{k}\cdot 
\mathbf{a}_{1}}$, $a_{16}=a_{17}=a_{18}=0$, $a_{19}=\lambda +\lambda ^{\ast
}e^{-ik_{x}}$, $a_{21}=a_{22}=a_{23}=\sqrt{t}$, $a_{24}=a_{25}=0$, 
$a_{26}=\lambda ^{\ast }+\lambda e^{i\mathbf{k}\cdot \mathbf{a}_{1}}$, 
$a_{27}=\lambda +\lambda ^{\ast }e^{-i\mathbf{k}\cdot \mathbf{a}_{2}}$, 
$a_{28}=\lambda ^{\ast }+\lambda e^{-ik_{x}}$, $a_{29}=0$ and 
$\mathbf{a}_{1}=\left\{ \sqrt{3}/2,1/2\right\} $, $\mathbf{a}_{2}=\left\{ \sqrt{3}/2,-1/2\right\} $. 
The parent Hamiltonian $H_{\text{par}}$ is found to be 
\begin{equation}
H_{\text{par}}=3\left( \left\vert t\right\vert +2\left\vert \lambda
\right\vert \right) +H_{\text{Haldane}}.
\end{equation}%
The chiral edge state in nanoribbon geometry emerges in $H_{\text{root}}$,
as shown in Fig.\ref{FigSqHaldane}(d). Furthermore, there are $7$ zero
energy states in $H_{\text{total}}$ due to the Lieb theorem\cite{Lieb} with $\left\vert M-N\right\vert =7$.

\begin{figure}[t]
\centerline{\includegraphics[width=0.49\textwidth]{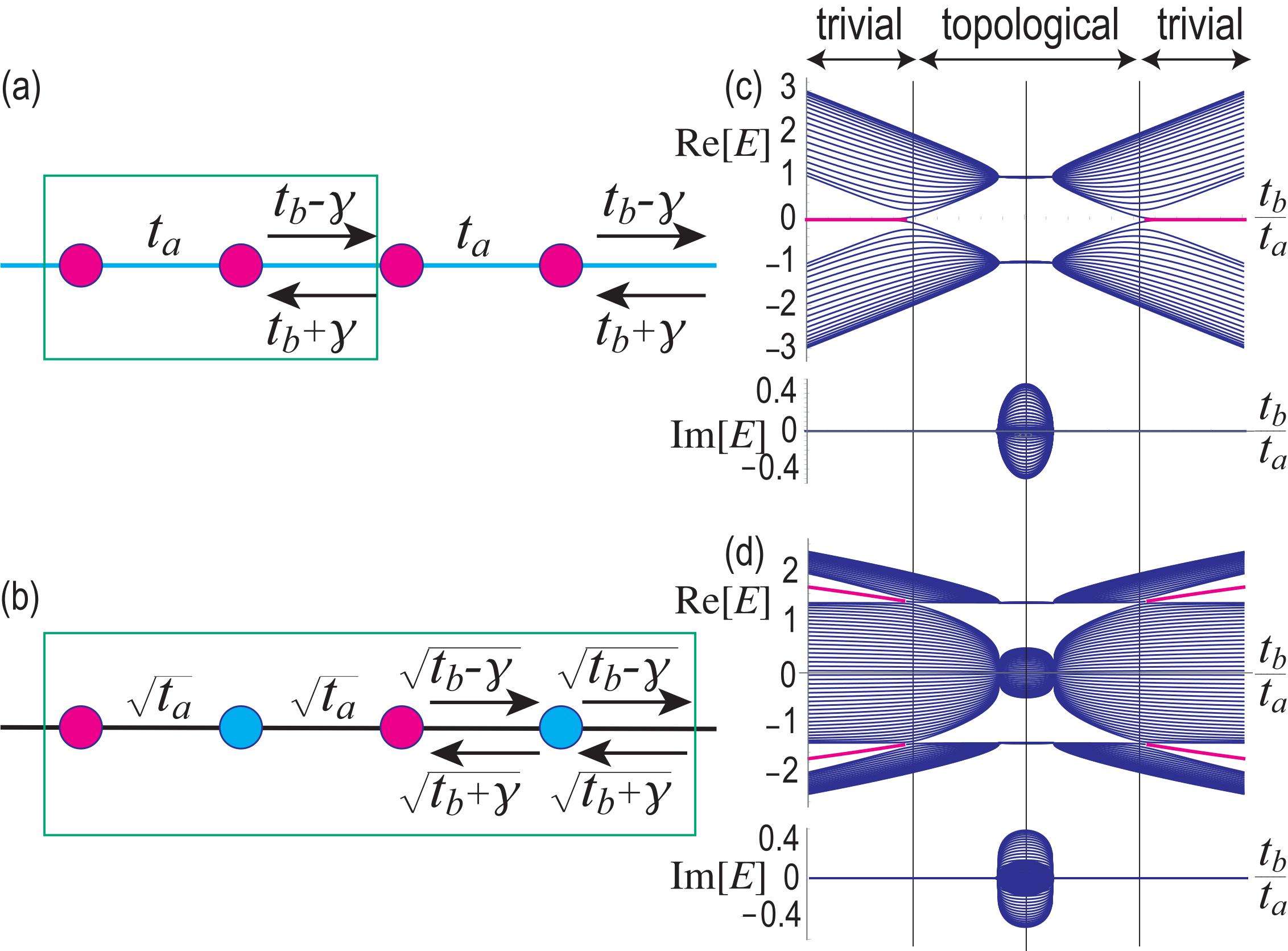}}
\caption{Illustration of (a) the graph and (b) the subdivided graph for the
nonreciprocal non-Hermitian SSH model $H_{\text{SSH}}^{\text{non}}$. (c)
Real and imaginary parts of the energy spectrum of $H_{\text{SSH}}^{\text{non}}$ 
and (d) $H_{\text{root}}$ in unit of $t_a$ as a function of $t_{b}/t_{a}$. 
The topological edge states are marked in magenta. In-gap states are on
the curve $E=\pm \protect\sqrt{\left\vert t_{a}\right\vert +\protect\sqrt{t_{b}^{2}-\protect\gamma ^{2}}}$ 
for $H_{\text{root}}$. We have set $\protect\gamma =t_a/4$.}
\label{FigNonrecipro}
\end{figure}

\textbf{Square-root non-Hermitian SSH model:} We proceed to construct a
square root of a non-Hermitian SSH model by introducing the nonreciprocity 
$\gamma $, as illustrated in Fig.\ref{FigNonrecipro}(a). The Hamiltonian reads%
\cite{Scho,Lieu,Lee,Yin,Yao,EzawaSSH}%
\begin{equation}
H_{\text{SSH}}^{\text{non}}=\left( 
\begin{array}{cc}
0 & t_{a}+\left( t_{b}+\gamma \right) e^{-ik} \\ 
t_{a}+\left( t_{b}-\gamma \right) e^{ik} & 0%
\end{array}%
\right) ,
\end{equation}%
where the hopping amplitudes are different between left and right goings.
The spectrum contains zero-energy edges states in the topological phase,
whose real and imaginary parts are shown in Fig.\ref{FigNonrecipro}(c) and
(c'). The square-root Hamiltonian $H_{\text{root}}$ is defined on the
subdivided graph in Fig.\ref{FigNonrecipro}(b), and given by the Hamiltonian
(\ref{HamilSR}) with%
\begin{align}
H_{2\times 2}^{\text{left}} &=\left( 
\begin{array}{cc}
t_{a} & \sqrt{t_{b}+\gamma }e^{-ik} \\ 
t_{a} & \sqrt{t_{b}-\gamma }%
\end{array}%
\right) , \\
H_{2\times 2}^{\text{right}} &=\left( 
\begin{array}{cc}
t_{a} & t_{a} \\ 
\sqrt{t_{b}-\gamma }e^{ik} & \sqrt{t_{b}+\gamma }%
\end{array}%
\right) .
\end{align}%
The parent Hamiltonian $H_{\text{par}}$ is found to be%
\begin{equation}
H_{\text{par}}=\left\vert t_{a}\right\vert +\sqrt{t_{b}^{2}-\gamma ^{2}}+H_{\text{SSH}}^{\text{non}}.
\end{equation}%
The residual Hamiltonian is given by $H_{\text{res}}=\{a_{ij}\}$, where 
$a_{11}=2t_{a}$, $a_{12}=\sqrt{t_{a}}\left( \sqrt{t_{b}-\gamma }+\sqrt{t_{b}+\gamma }e^{ik}\right) $, 
$a_{21}=\sqrt{t_{a}}\left( \sqrt{t_{b}+\gamma }+\sqrt{t_{b}-\gamma }e^{ik}\right) $, 
$a_{22}=2\sqrt{t_{b}^{2}-\gamma ^{2}}$. 
In-gap edge states emerge at $E=\pm \sqrt{\left\vert t_{a}\right\vert +%
\sqrt{t_{b}^{2}-\gamma ^{2}}}$ for $\left\vert t_{a}\right\vert >\left\vert
t_{b}\right\vert $ in $H_{\text{root}}$, as shown in Fig.\ref{FigNonrecipro}(d), 
which are transformed from the zero-energy topological edge states in $H_{\text{SSH}}^{\text{non}}$.

\textbf{Square-root non-Hermitian topological insulator: }In general, we
obtain a square root of a non-Hermitian topological system by taking (\ref%
{HamilSR}) with $H_{M\times N}^{\text{right}}\neq \left( H_{N\times M}^{\text{left}}\right) ^{\dagger }$. For example, we take 
\begin{equation}
H_{2\times 2}^{\text{left}}=\left( 
\begin{array}{cc}
t_{1}^{\text{left}} & t_{4}^{\text{left}}e^{-ik} \\ 
t_{2}^{\text{left}} & t_{3}^{\text{left}}%
\end{array}%
\right) ,\quad H_{2\times 2}^{\text{right}}=\left( 
\begin{array}{cc}
t_{1}^{\text{right}} & t_{2}^{\text{right}} \\ 
t_{4}^{\text{right}}e^{ik} & t_{3}^{\text{right}}%
\end{array}%
\right) .
\end{equation}%
By calculating $(H_{\text{root}})^{2}=H_{\text{par}}\oplus H_{\text{res}}$,
we obtain%
\begin{equation}
H_{\text{par}}=\left( 
\begin{array}{cc}
t_{1}^{\text{left}}t_{1}^{\text{right}}+t_{4}^{\text{left}}t_{4}^{\text{right%
}} & t_{1}^{\text{left}}t_{2}^{\text{right}}+t_{4}^{\text{left}}t_{3}^{\text{%
right}}e^{-ik} \\ 
t_{2}^{\text{left}}t_{1}^{\text{right}}+t_{3}^{\text{left}}t_{4}^{\text{right%
}}e^{ik} & t_{2}^{\text{left}}t_{2}^{\text{right}}+t_{3}^{\text{left}}t_{3}^{%
\text{right}}%
\end{array}%
\right) ,
\end{equation}%
which is nonreciprocal non-Hermitian in general.

\textbf{Discussions:} We have presented a systematic method to construct
square-root topological insulators and superconductors based on subdivided
graphs. We recall that subdivided graphs naturally arise in
electric-circuits when we rewrite the Kirchhoff law into the Schr\"{o}dinger
equation\cite{EzawaSch,EzawaUniv}. Hence, it would be natural to make
experimental observation of square-root topological systems with the use of
electric circuits. We start with a lattice electric circuit. In the original
graph, it contains voltage at the sites, which correspond to the vertices in
the graph theory. We can define currents flowing between two adjacent sites,
which corresponds to links in the graph theory. Both the in-gap
nonzero-energy edge states and the zero-energy flat bands due to the Lieb
theorem are to be observed by measuring impedance peaks\cite%
{TECNature,ComPhys,EzawaTEC}. Another possibility to realize square-root
topological systems is a direct construction of lattice structures by
photonic\cite{Kremer} or acoustic systems\cite{Xue,Khani}.

\clearpage\newpage \onecolumngrid

\setcounter{section}{0} \setcounter{figure}{0} \setcounter{equation}{0}

\centerline{\textbf{\Large Supplemental Material}}\bigskip 

\centerline{\large\textbf{Systematic construction of square-root topological
insulators and superconductors}} \medskip \centerline{Motohiko Ezawa} 
\centerline{Department of
Applied Physics, University of Tokyo, Hongo 7-3-1, 113-8656, Japan}

\section{Naive construction of a square root of a Hamiltonian}

We try to construct a square-root of a given Hamiltonian in a naive way,
where we take a square root of a matrix representing the original
Hamiltonian. First, we diagonalize the original Hamiltonian $H$ by a unitary
transformation as%
\begin{equation}
U^{-1}HU=H_{\text{D}},
\end{equation}%
where 
\begin{equation}
H_{\text{D}}=\text{diag}.\left( \varepsilon _{1},\cdots ,\varepsilon
_{N}\right)
\end{equation}%
is a diagonal matrix whose components are eigenvalues $\varepsilon _{j}$
with $1\leq j\leq N$ being $N$ a dimension of the matrix $H$ and $H_{\text{D}%
}$. Then a square-root Hamiltonian $\sqrt{H}$ is given by%
\begin{equation}
\sqrt{H}=U\sqrt{H_{\text{D}}}U^{-1},
\end{equation}%
where%
\begin{equation}
\sqrt{H_{\text{D}}}=\text{diag}.\left( \sqrt{\varepsilon _{1}},\cdots ,\sqrt{%
\varepsilon _{N}}\right) .
\end{equation}%
A problem is that a square-root Hamiltonian $\sqrt{H}$ is an infinite-range
hopping model even when we start with a local hopping model $H$. We see it
for an example of the square root of the Su-Schrieffer-Heeger model (\ref%
{HamilSSH}), or%
\begin{equation}
H_{\text{SSH}}=\left( 
\begin{array}{cc}
0 & t_{a}+t_{b}e^{-ik} \\ 
t_{a}+t_{b}e^{ik} & 0%
\end{array}%
\right) .
\end{equation}%
It is diagonalize as%
\begin{equation}
H_{\text{D}}=E\left( k\right) \sigma _{z}
\end{equation}%
with an energy%
\begin{equation}
E\left( k\right) =\sqrt{t_{a}^{2}+t_{b}^{2}+2t_{a}t_{b}\cos k},
\end{equation}%
and a unitary matrix%
\begin{equation}
U=\frac{1}{\sqrt{2}}\left( 
\begin{array}{cc}
\frac{E\left( k\right) }{t_{a}+t_{b}e^{-ik}} & \frac{-E\left( k\right) }{%
t_{a}+t_{b}e^{-ik}} \\ 
1 & 1%
\end{array}%
\right) .
\end{equation}%
Then the square-root Hamiltonian is given by%
\begin{equation}
\sqrt{H}=\left( 
\begin{array}{cc}
0 & \frac{\sqrt{E\left( k\right) }}{t_{a}+t_{b}e^{-ik}} \\ 
\frac{t_{a}+t_{b}e^{-ik}}{\sqrt{E\left( k\right) }} & 0%
\end{array}%
\right) ,
\end{equation}%
which is an infinite-range hopping model.

\section{Bipartite graph}

We have constructed the Hamiltonian $H_{\text{root}}$ on the subdivided
graph and decomposed it as $(H_{\text{root}})^{2}=H_{\text{par}}\oplus H_{%
\text{res}}$. The eigenvalues of $H_{\text{par}}$ and $H_{\text{res}}$ have
the following properties.

1) All of the eigenvalues are identical between $\varepsilon _{\text{par}%
}=\varepsilon _{\text{res}}$ except for the zero energy.

2) All of the eigenvalues are non-negative $\varepsilon _{\text{par}}\geq 0$
and $\varepsilon _{\text{res}}\geq 0$.

\noindent {}Let us prove them.

1) We study the eigen equation%
\begin{equation}
H_{\text{par}}\left\vert \psi _{\text{par}}\right\rangle =H_{N\times M}^{%
\text{left}}H_{M\times N}^{\text{right}}\left\vert \psi _{\text{par}%
}\right\rangle =\varepsilon \left\vert \psi _{\text{par}}\right\rangle
\end{equation}%
with $\varepsilon \neq 0$. We multiply $H_{M\times N}^{\text{right}}$ from
the left and obtain%
\begin{equation}
H_{M\times N}^{\text{right}}H_{N\times M}^{\text{left}}\left( H_{M\times N}^{%
\text{right}}\left\vert \psi _{\text{par}}\right\rangle \right) =\varepsilon
H_{M\times N}^{\text{right}}\left\vert \psi _{\text{par}}\right\rangle .
\end{equation}%
By defining 
\begin{equation}
\left\vert \psi _{\text{res}}\right\rangle \equiv H_{M\times N}^{\text{right}%
}\left\vert \psi _{\text{par}}\right\rangle ,
\end{equation}%
we obtain%
\begin{equation}
H_{\text{res}}\left\vert \psi _{\text{res}}\right\rangle =\varepsilon
\left\vert \psi _{\text{res}}\right\rangle .
\end{equation}%
Hence the eigenvalues are identical between $H_{\text{par}}$ and $H_{\text{%
res}}$.

2) When $H_{\text{root}}$ is Hermitian, it is necessary that%
\begin{equation}
H_{\text{par}}=(H_{M\times N}^{\text{right}})^{\dagger }H_{M\times N}^{\text{%
right}},\qquad H_{\text{res}}=(H_{N\times M}^{\text{left}})^{\dagger
}H_{N\times M}^{\text{left}}.
\end{equation}%
For $N$-dimensional vector $\psi _{N}$ and $M$-dimensional vector $\psi _{M}$%
, we find%
\begin{align}
\langle \psi _{N},H_{\text{par}}\psi _{N}\rangle &=\langle H_{M\times N}^{%
\text{right}}\psi _{N},H_{M\times N}^{\text{right}}\psi _{N}\rangle
=|H_{M\times N}^{\text{right}}\psi _{N}|^{2}\geq 0, \\
\langle \psi _{M},H_{\text{res}}\psi _{M}\rangle &=\langle H_{N\times M}^{%
\text{left}}\psi _{M},H_{N\times M}^{\text{left}}\psi _{M}\rangle
=|H_{N\times M}^{\text{left}}\psi _{M}|^{2}\geq 0,
\end{align}%
implying $\varepsilon _{\text{par}}\geq 0$ and $\varepsilon _{\text{res}%
}\geq 0$.

\end{document}